# Josephson diode effect in nanowire-based Andreev molecules

Shang Zhu[1,2,#], Yiwen Ma[1,2,#], Jiangbo He[1,3], Xiaozhou Yang[1,2], Zhongmou Jia[1,2], Min Wei[1,2], Yiping Jiao[1,4], Jiezhong He[1,2], Enna Zhuo[1,2], Xuewei Cao[4], Bingbing Tong[1,5], Ziwei Dou[1], Peiling Li[1,5], Jie Shen[1], Xiaohui Song[1,5], Zhaozheng Lyu[1,5], Guangtong Liu[1,5], Dong Pan[6,*], Jianhua Zhao[6,7], Bo Lu[8,*], Li Lu[1,2,5,*], Fanming Qu[1,2,5,*]

[1] Beijing National Laboratory for Condensed Matter Physics, Institute of Physics, Chinese Academy of Sciences, Beijing 100190, China

[2] University of Chinese Academy of Sciences, Beijing 100049, China

[3] Key Laboratory of Low-Dimensional Quantum Structures and Quantum Control of Ministry of Education, Department of Physics and Synergetic Innovation Center of Quantum Effects and Applications, Hunan Normal University, Changsha 410081, China

[4] School of Physics, Nankai University, Tianjin 300071, China

[5] Hefei National Laboratory, Hefei 230088, China

[6] State Key Laboratory of Superlattices and Microstructures, Institute of Semiconductors, Chinese Academy of Sciences, Beijing 100083, China

[7] National Key Laboratory of Spintronics, Hangzhou International Innovation Institute, Beihang University, Hangzhou 311115, China

[8] Department of Physics, Tianjin University, Tianjin 300072, China

[#] These authors contributed equally to this work.

[*] Email: pandong@semi.ac.cn; billmarx@tju.edu.cn; lilu@iphy.ac.cn; fanmingqu@iphy.ac.cn

**Superconducting systems exhibit non-reciprocal current transport under certain conditions of symmetry breaking, a phenomenon known as the superconducting diode effect. This effect allows for perfect rectification of supercurrent, and has received considerable research interest. We report the observation of the Josephson diode effect (JDE) in nanowire-based Andreev molecules, where the time-reversal and spatial-inversion symmetries of a Josephson junction (JJ) can be nonlocally broken by coherently coupling to another JJ. The JDE can be controlled using both non-local phase and gate voltages. Notably, the non-local phase can induce a sign reversal of the**

**diode efficiency, a manifestation of regulating the probabilities of double elastic cotunneling and double-crossed Andreev reflection. Additionally, the diode efficiency can be further modulated by local and non-local gate voltages, exhibiting a central-peak feature in the gate-voltage space. Our theoretical calculations of the energy spectrum and the Josephson currents align well with the experimental results. These results demonstrate the non-local regulation of the JDE in Andreev molecules, offering significant implications for the control of multi-JJ devices and the development of advanced superconducting devices.**

## Introduction

The superconducting diode effect (SDE) exhibits direction-dependent switching current, making it a promising candidate for a dissipationless circuit element, with potential applications in superconducting technologies. SDE typically requires the breaking of both time-reversal and spatial-inversion symmetries in superconducting devices[1]. So far, the SDE has been realized in various platforms, including superconducting films[2-15] and Josephson junctions (JJs)[16-41], with the latter known as the Josephson diode effect (JDE). Moreover, a variety of studies have explored the origins of symmetry breaking, including magnetic interactions of internal magnetic layers or external magnetic fields[2-4,17,22], vortices[6,7,9,10,18], higher harmonics in the current-phase relation (CPR)[20,25,38,39] and finite-momentum pairing[15,21,30,33]. The JDE has also been observed in devices comprising multiple JJs, such as SQUID devices[36-39], multiterminal devices[25,41], and coherently coupled two JJs[28,40].

One particular scenario is the JDE in Andreev molecules where *non-local* control is enabled. An Andreev molecule is composed of two closely spaced JJs, and it has been theorized that when the distance $l$ between the two JJs is smaller than the superconducting coherence length $\xi_0$, their Andreev spectra hybridize into a molecular state[42]. The hybridization of Andreev bound states (ABSs) results in an avoided crossing structure, and these ABSs can be pushed out of the energy gap into the continuum states. The supercurrent of one JJ can be regulated non-locally by the phase of the other JJ, leading to a non-local Josephson effect[42-45] and the JDE[46]. The origin of the JDE is attributed to the asymmetric ABSs energy spectrum and the continuum states.

At the microscopic level, symmetry breaking in Andreev molecules originates from the phase competition between double elastic cotunneling (dECT) and double-crossed Andreev reflection (dCAR) of Cooper pairs[46], a mechanism that is rarely discussed in studies of the JDE[1]. In normal-superconductor-normal (N-S-N) devices, elastic cotunneling (ECT) and crossed Andreev reflection (CAR) are the essential mechanisms in the construction of artificial Kitaev chains[47-49] with poor man's Majorana zero modes[50-54]. In Andreev molecules, which can be viewed as a configuration of S-N-S-N-S, the JDE is closely related to the relative probabilities of dECT and dCAR, which in turn depend on the non-local phase[46].

In this work, we report the observation of the JDE in Andreev molecules constructed from InAs nanowires. By varying both non-local phase and gate voltages, we demonstrate the control over the JDE. Sign reversal of the diode efficiency can be achieved by tuning the non-local phase, demonstrating the regulation of the dECT and dCAR processes. In the local and non-local gate-voltage space, the diode efficiency exhibits a central-peak structure. Our theoretical calculations based on Furusaki-Tsukada's method and exact diagonalization method show consistent behavior of the Andreev molecules. Our results present the non-local regulation ability of the JDE in Andreev molecules by phase and gates, paving the way for further exploration of two-JJ coupling mechanisms and the non-local modulation of multi-JJ devices.

## Results

**Origin of JDE in an Andreev molecule.** High-quality InAs nanowires with an in-situ epitaxial half shell of 15 nm-thick Al are grown by molecular beam epitaxy[55]. Andreev molecule devices are fabricated using standard electron beam lithography and high-precision transfer techniques on selected nanowires [Figs. 1a and 1b]. Two short JJs are formed by wet etching of Al, each approximately 50 nm in length, ensuring a large supercurrent of ~11 nA. A thin Al strip with a length ($l$) of about 200 nm is retained between the two JJs to enable coherent coupling for the formation of an Andreev molecule. An additional Al layer is deposited to form superconducting leads, as well as a superconducting loop on the right JJ to control its phase. Figure 1a is a close-up of the full device shown in Fig. 1b. All experimental data presented in the main text are mainly measured on device A in a dilution refrigerator at about 10 mK using standard lock-in techniques. Other data from device B are presented in the Supplementary Information (see Supplementary Notes 5, 6 and 7).

The device of an Andreev molecule can be represented by the schematics in Fig. 1c. The phase differences of the JJs on the left and right are denoted as $\delta_L$ and $\delta_R$, respectively, and $\delta_R$ can be controlled by magnetic flux through a superconducting loop. A current drive $I$ is applied from the left superconducting electrode, and the middle superconductor is grounded. The gate voltages $V_{GL}$ and $V_{GR}$ modulate the supercurrent (chemical potential) of the two JJs, respectively. In this configuration of electrical measurement on the left JJ, $V_{GL}$ and $V_{GR}$ are the local and non-local gate voltages, and $\delta_L$ and $\delta_R$ are the local and non-local phase differences, respectively. When the distance $l$ between the two JJs is shorter than the superconducting coherence length $\xi_0$, the Andreev wave functions hybridize, resulting in an Andreev molecule with a typical energy spectrum shown in Fig. 1d, which is obtained by the exact diagonalization method for a fixed $\delta_R = 0.85\pi$ (see Supplementary Note 2). For energies $|E| < \Delta$, the Andreev spectrum exhibits avoided crossings at the degenerate points due to coherent coupling, with the color of the lines indicating the supercurrent direction carried by ABSs, red for positive and blue for negative. The ABSs can be expelled into the continuum states with $|E| \geq \Delta$, forming Andreev scattering states[46], whose supercurrent density $j_L$ is indicated by the color. The calculation of $I_L$ (the supercurrent of the left JJ) is carried out by considering the contribution from both discrete sub-gap ABSs and the continuum states outside the gap. At zero temperature, $I_L$ is given by[46]:

$$I_L = \frac{1}{\Phi_0} \Sigma_{E_{ABS}<0} \frac{\partial E_{ABS}}{\partial \delta_L} + I_L^{cont.}. \tag{1}$$

The first term stands for the contribution of the negative-energy ABSs, with $\Phi_0 = h/2e$, where $h$ is the Planck's constant and e is the elementary charge. The second term accounts for the contribution from the continuum states with negative energy.

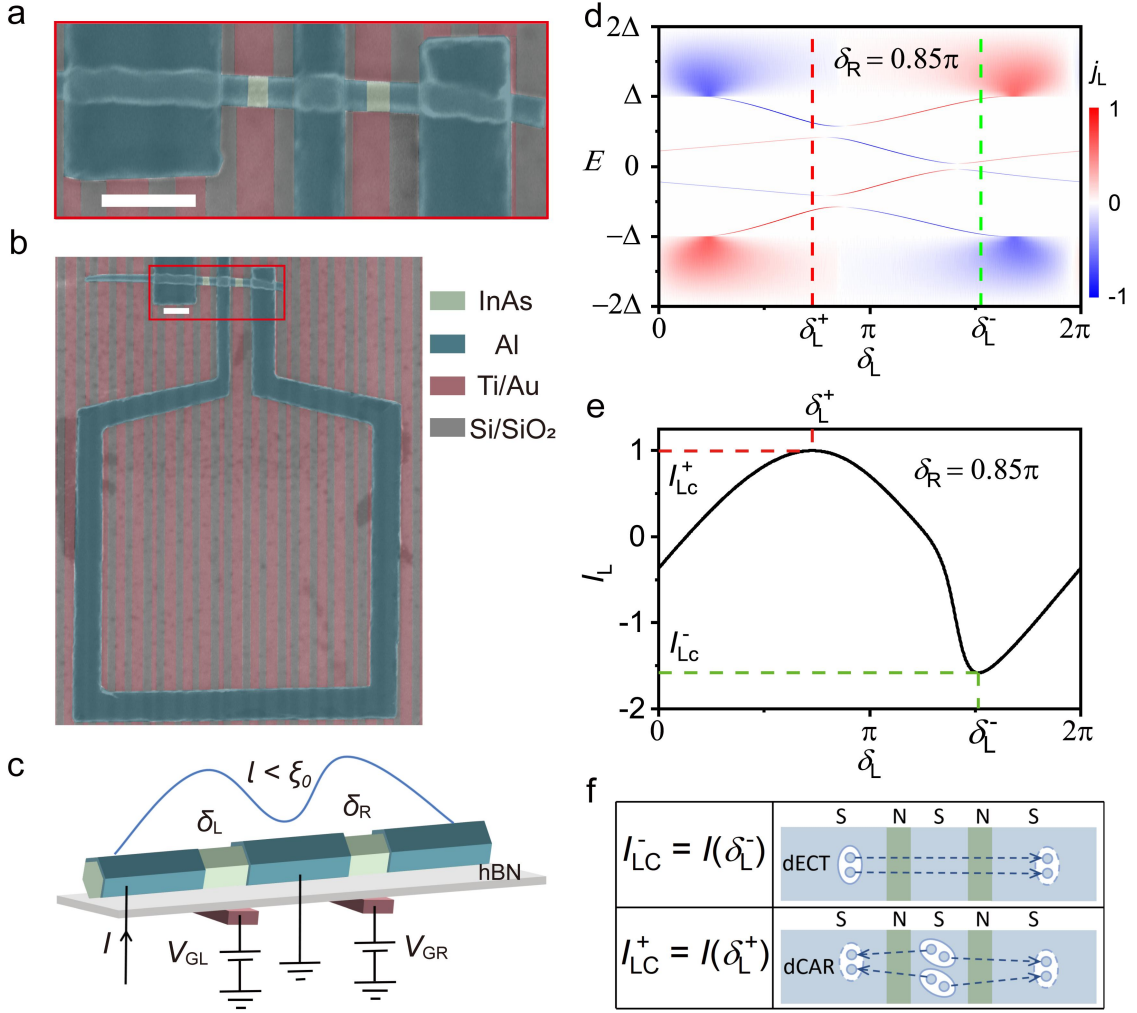

**Fig. 1. Device of an Andreev molecule. a, b** False-color scanning electron microscope (SEM) images of device A. The scale bar is 200 nm. The hBN dielectric layer sits between the Ti/Au finger gates and the nanowire, but is transparent in this image. **c** Schematics of an Andreev molecule. The separation $l$ of the two JJs is less than the superconducting coherence length $\xi_0$. $\delta_{L,R}$ are the phase differences of the left and right JJ, respectively. The solid blue line represents the overlap of the Andreev wave functions in the two JJs. **d** Energy spectrum of the Andreev molecule at $\delta_R = 0.85\pi$ calculated using the exact diagonalization method. For $|E| < \Delta$, the red and blue color of the lines indicate the positive and negative supercurrent of the left JJ carried by the ABSs, respectively. For $|E| \geq \Delta$, the color corresponds to the supercurrent density $j_L$ of the continuum states, as indicated by the color bar. **e** CPR of the left JJ at $\delta_R = 0.85\pi$ calculated using the exact diagonalization method. The red and green dashed lines in (d, e) mark the positions of $I_{Lc}^+$ at $\delta_L = \delta_L^+$ and $I_{Lc}^-$ at $\delta_L = \delta_L^-$, respectively. **f** Illustration of the dECT and dCAR processes at $\delta_R = 0.85\pi$.

From Eq. 1, we can obtain the CPR of the left JJ. When $\delta_R = 0 \pmod{\pi}$, the CPR is symmetric, and the positive and negative critical supercurrents are equal in amplitude, without JDE. However, the CPR is asymmetric when $\delta_R \neq 0 \pmod{\pi}$[46] and the JDE appears. For example, Fig. 1e displays the asymmetric CPR at $\delta_R = 0.85\pi$, as well as the characteristic of a $\varphi_0$ junction. The positive critical supercurrent $I_{Lc}^+ = I_L(\delta_L^+)$ is not equal to (smaller than) the absolute value of the negative critical supercurrent $|I_{Lc}^-| = |I_L(\delta_L^-)|$, thus demonstrating the presence of the JDE.

Microscopically, the coupling between the two JJs arises primarily from the dECT and dCAR processes, as illustrated in Fig. 1f. In the dECT process, a Cooper pair in the left superconductor crosses the middle superconductor to the right superconductor, with $I_L$ and $I_R$ propagating in the same direction, where $I_R$ is the supercurrent of the right JJ. In contrast, for the dCAR process, two Cooper pairs in the middle superconductor split and form new Cooper pairs in the left and right superconductors, with $I_L$ and $I_R$ propagating in opposite directions. As shown in Fig. 1d, at $\delta_L = \delta_L^-$, both the ABSs and the continuum states (at $E < 0$) contribute the same-direction supercurrent to $I_{Lc}^-$, where dECT is the dominant mechanism. In contrast, at $\delta_L = \delta_L^+$, the supercurrent direction carried by the two ABSs is opposite, i.e., dCAR is dominant, resulting in a largely reduced critical supercurrent $I_{Lc}^+$. In addition, the contribution of the continuum states is also smaller at $\delta_L = \delta_L^+$ than at $\delta_L = \delta_L^-$, further enhancing the JDE. These results, as an example at $\delta_R = 0.85\pi$, indicate that the critical supercurrent $|I_{Lc}^-|$ when dECT dominates is greater than $I_{Lc}^+$ when dCAR dominates, consistent with the asymmetric CPR shown in Fig. 1e.

Importantly, by tuning the non-local phase $\delta_R$, such scenario can be reversed (see Supplementary Figs. S2a, b). At $\delta_R = 1.15\pi$, the critical supercurrent $|I_{Lc}^-|$ when dCAR dominates is less than $I_{Lc}^+$ when dECT dominates. Performing additional calculations at varying $\delta_R$ (see Supplementary Figs. S3, S4), when $0 < \delta_R < \pi$, $I_{Lc}^+ < |I_{Lc}^-|$, dECT dominates at $\delta_L = \delta_L^-$; while $\pi < \delta_R < 2\pi$, $I_{Lc}^+ > |I_{Lc}^-|$, dECT dominates at $\delta_L = \delta_L^+$. The critical supercurrent in the dECT-dominated regime always exceeds that in the dCAR-dominated regime. As the phase is tuned, the $|I_{Lc}^-|$ transitions from dECT-dominated to dCAR-dominated behavior, while the $I_{Lc}^+$ exhibits the opposite trend, resulting in the reversal of the direction of rectification.

**Non-local phase control of JDE.** We next present the experimental results. Figure 2a shows the differential resistance $dV/dI$ of the left JJ as a function of $I$ and perpendicular magnetic field $B$ with $V_{GL} = V_{GR} = 0$ V. The positive and negative halves of Fig. 2a are respectively scanned from $I = 0$ to the positive and negative currents, as indicated by the red and green arrows. As a function of $B$, which controls $\delta_R$, the local positive and negative critical supercurrent show periodic oscillations, demonstrating the non-local Josephson effect[42-45]. The red and green dashed lines mark the maximum $I_{Lc}^+$ and $|I_{Lc}^-|$, respectively, which occur at different $B$. $I_{Lc}^+$ and $|I_{Lc}^-|$ can be extracted from Fig. 2a, as plotted in Fig. 2b. It can be seen that within one period, $I_{Lc}^+$ and $|I_{Lc}^-|$ intersect twice, while under other magnetic fields $I_{Lc}^+ \neq |I_{Lc}^-|$, which is the JDE. Noteworthy, the tiny magnetic field is used solely to regulate the phase of the right JJ rather than interact with SOC[2,3,17,22], although InAs nanowires have strong SOC.

In order to characterize the JDE, the diode efficiency is typically defined as $\eta = (I_{Lc}^+ - |I_{Lc}^-|)/(I_{Lc}^+ + |I_{Lc}^-|)$, as shown in Fig. 2c. $\eta$ varies periodically with $B$, and a sign reversal occurs whenever $\eta$ crosses zero. As shown in Figs. 1d-f and Figs. S2a, b, at $\delta_R = 0.85\pi$, the dominant mechanisms are dCAR and dECT for $I_{Lc}^+$ and $I_{Lc}^-$, respectively, and $I_{Lc}^+ < |I_{Lc}^-| \rightarrow \eta < 0$. On the contrary, when $\delta_R = 1.15\pi$, the dominant mechanism is reversed, and $I_{Lc}^+ > |I_{Lc}^-| \rightarrow \eta > 0$. Therefore, the sign reversal of $\eta$ demonstrates that non-local phase control can modulate the probabilities of dECT and dCAR in Andreev molecules.

We find that the maximum $\eta$ is approximately 2.3%, which is much lower than the highest theoretical prediction of 45%[46]. We attribute this discrepancy to the low junction transmission, which significantly suppresses the diode efficiency. By fitting the data using the Octavio-Tinkham-Blonder-Klapwijk (OTBK) theory, the junction transmission $\tau \approx 0.53$ is extracted for $V_{GL} = V_{GR} = 0$ V (see Supplementary Note 4), which is much lower than the transmission $\tau = 1$ where the maximum efficiency is predicted. Ref. [43] also illustrated that $\eta$ remains below 2% for $\tau < 0.5$, and a high transmission would improve the efficiency significantly.

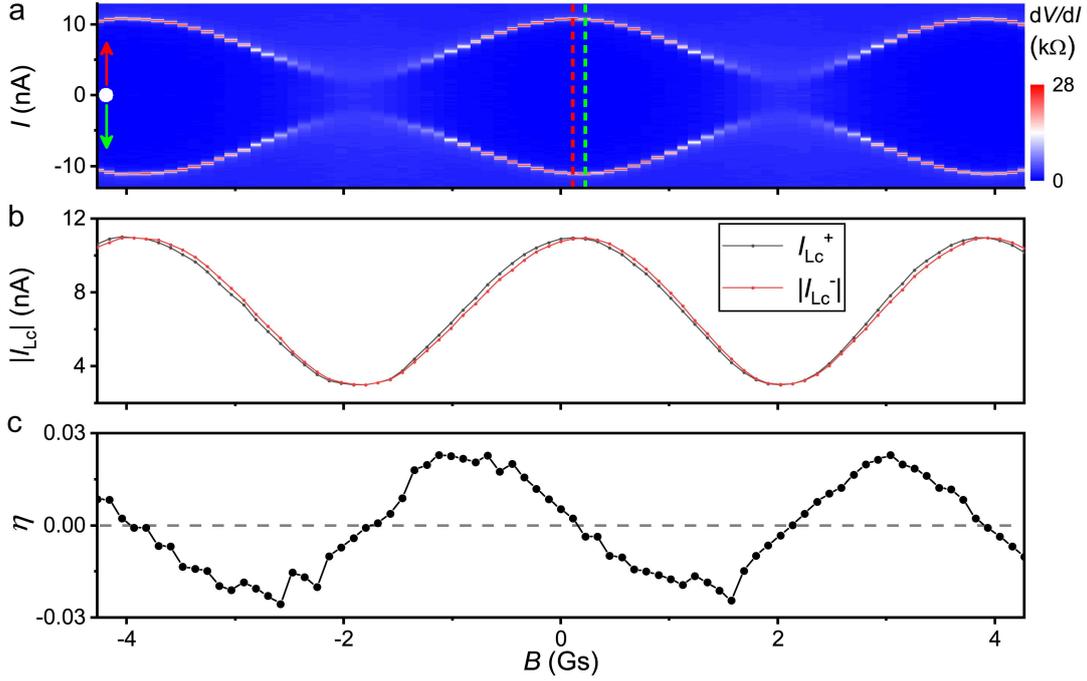

**Fig. 2. Josephson diode effect. a** The differential resistance $dV/dI$ of the left JJ as a function of $I$ and $B$ with $V_{GL} = V_{GR} = 0$ V. The red and green dashed lines indicate the maximum $I_{Lc}^+$ and $|I_{Lc}^-|$, respectively. The red and green arrows represent the scanning directions of the positive and negative current. **b** $|I_{Lc}|$ as a function of $B$. **c** Diode efficiency $\eta$ as a function of $B$.

**Local and non-local gates control of JDE.** We next demonstrate the tuning of the JDE by the local and non-local gates. Figure 3a shows the differential resistance $dV/dI$ of the left JJ as a function of $I$ and $V_{GL}$. The critical supercurrent decreases as $V_{GL}$ decreases, and completely disappears when $V_{GL} = -3.7$ V. Hence, the local gate applied to the left JJ can control the supercurrent in this junction. While the right JJ cannot be characterized separately due to the presence of a superconducting loop, a similar behavior is expected.

We then investigate the influence of the individual local ($V_{GL}$) or non-local ($V_{GR}$) gate on the JDE. Figure 3b presents $I_{Lc}^+$ and $|I_{Lc}^-|$ as a function of $B$ at several different $V_{GL}$, with a fixed $V_{GR} = -1.6$ V. The JDE can be observed at all these settings, although the critical supercurrent decreases significantly when $V_{GL} = -2.4$ V. Plots of $\eta$ vs. $B$ are displayed in Fig. 3d, illustrating the tuning of $\eta$ by the non-local phase. Around the maximum $\eta$ at a fixed $B = -1.012$ Gs, as indicated by the red dashed line, $\eta$ vs. $V_{GL}$ can be extracted, as shown in the inset. $\eta$ presents a peaked feature around $V_{GL} = -1.6$ V, decreasing whether $V_{GL}$ is increased or decreased. An identical behavior is observed when switching the role of the local ($V_{GL}$) and non-local ($V_{GR}$) gates in the measurement, as shown in Figs. 3c and e. $\eta$ also presents a peaked feature when varying $V_{GR}$ at $B = -1.012$ Gs, as shown by the inset of Fig. 3e. These results manifest that the JDE can be regulated by both the local and non-local gates.

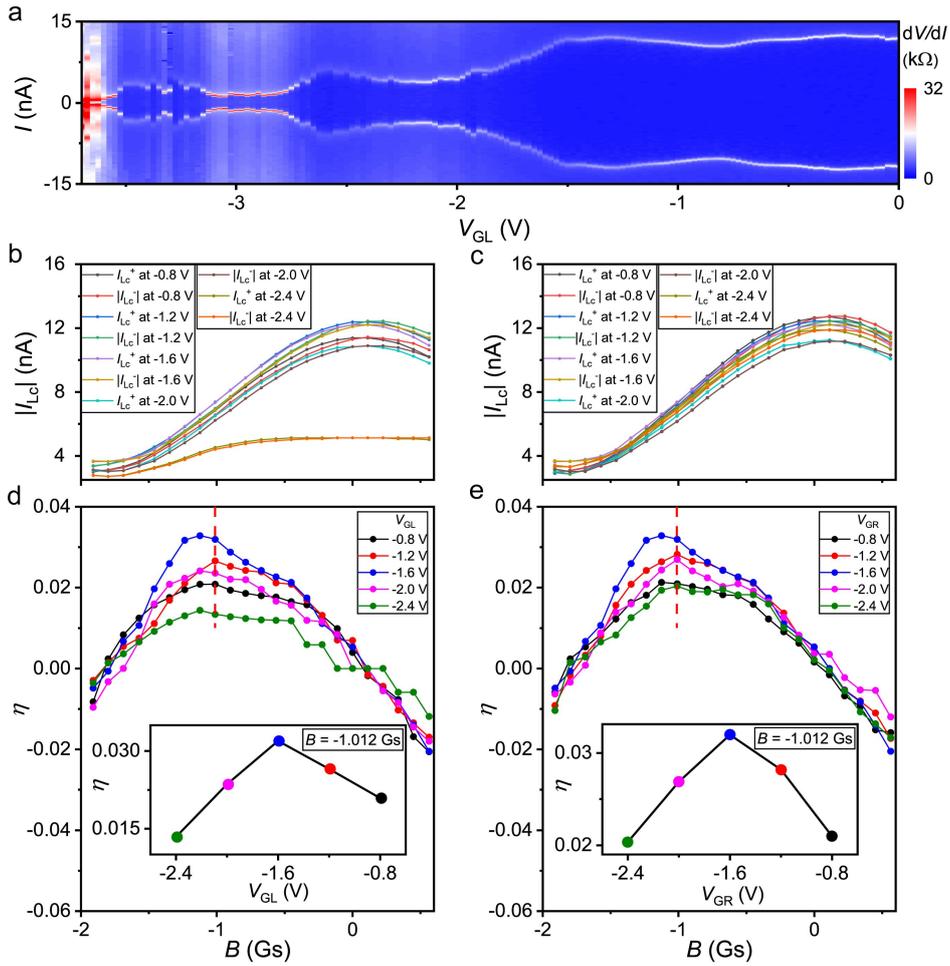

**Fig. 3. Dependence of JDE on individual gate. a** $dV/dI$ of the left JJ as a function of $I$ and $V_{GL}$. **b** $I_{Lc}^+$ and $|I_{Lc}^-|$ as a function of $B$ at $V_{GL} = -0.8, -1.2, -1.6, -2$ and $-2.4$ V, with a fixed $V_{GR} = -1.6$ V. **c** $I_{Lc}^+$ and $|I_{Lc}^-|$ as a function of $B$ at $V_{GR} = -0.8, -1.2, -1.6, -2$ and $-2.4$ V, with a fixed $V_{GL} = -1.6$ V. **d** $\eta$ vs. $B$ extracted from (b). The inset shows the $\eta$ vs. $V_{GL}$ curve at $B = -1.012$ Gs, as indicated by the red dashed line. **e** $\eta$ vs. $B$ extracted from (c). The inset shows the $\eta$ vs. $V_{GR}$ curve at $B = -1.012$ Gs, as indicated by the red dashed line.

In order to capture a full picture of the gate tuning on the JDE, we measured $\eta$ in the ($V_{GL}$, $V_{GR}$) two-dimensional space at $B = -1.012$ Gs, as shown in Fig. 4a. The green and red dashed lines correspond to the conditions in Figs. 3d and e, respectively. Near $V_{GL} = V_{GR} = -1.6$ V, $\eta$ is significantly higher than elsewhere, showing a central-peak feature. The horizontal lines extracted from Fig. 4a are plotted in Fig. 4b. Each individual line shows a peak, highlighted with larger dots. In fact, these larger dots are distributed along the black dashed line in Fig. 4a, which corresponds to the diagonal direction where $V_{GL} = V_{GR}$ and thus similar Josephson coupling in the two JJs. This indicates that the efficiency of the JDE is higher under more symmetric conditions.

Figures 4c and 4d are measured in device B. Within the explored gate-voltage space, the maximum $\eta$ is located near $V_{GL} = -1$ V and $V_{GR} = -0.75$ V. The larger dots in Fig. 4d are distributed along the black dashed line in Fig. 4c. In device B, we observed analogous diagonal distribution patterns, but the maximum $\eta$ lies outside the central region of the explored gate-voltage space. This is due to the limited gate-voltage range we can explore in device B, since when $V_{GL} < -1$ V, the critical supercurrent is too small to obtain reliable data (see Supplementary Note 7). Intrinsic device limitations restrict broader exploration, yet the acquired data exhibit behavior consistent with device A.

We now move to provide theoretical calculations of the diode efficiency. We use the Furusaki-Tsukada's method to calculate the Josephson current since it is efficient to take into account the complicated configurations of the gate voltages in our setup (see Supplementary Note 1). The effect of the gate voltages actually tunes the chemical potential ($U_{L,R}$) of the JJs, so in the model $V_{GL}$ and $V_{GR}$ are represented by $U_L$ and $U_R$, respectively. The calculated results are shown in Fig. 4e and Fig. 4f, in the form of two-dimensional and three-dimensional graphs, respectively. It can be seen that $\eta$ is highest near $U_L = U_R = 0$, presenting a central-peak feature. In addition, the peak of $\eta$ extends along the $U_L = U_R$ diagonal direction. These theoretical results are in good agreement with our experimental observations.

The underlying mechanism of such behavior might be primarily attributed to the wave-vector mismatch when $U_{L,R}$ and $U_S$ are different, where $U_S$ is the chemical potential of the superconducting regions and is set to zero in the model. At the center of Fig. 4e, $U_L = U_R = U_S$, and the wave-vector mismatch is the smallest, resulting in the largest transmission and maximum $\eta$. Along the diagonal where $U_L = U_R$, the symmetric condition of the two JJs also ensures a large $\eta$.

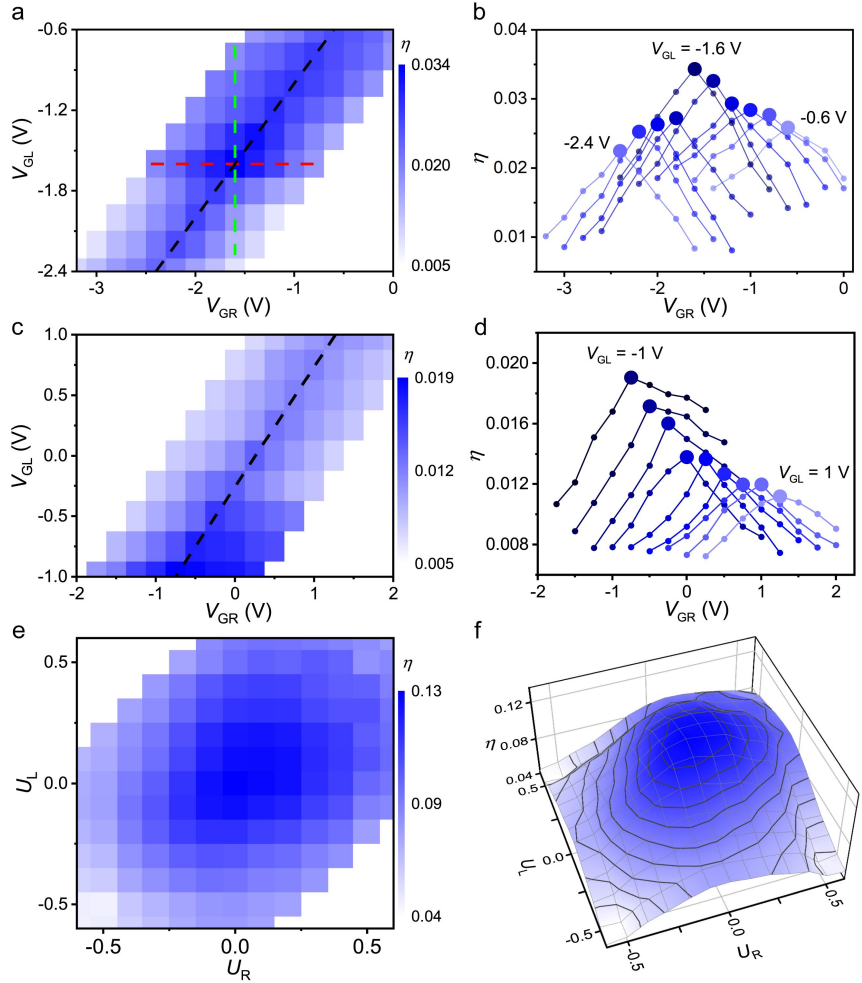

**Fig. 4. JDE in the local and non-local gate-voltage space. a** $\eta$ as a function of $V_{GL}$ and $V_{GR}$ at $B = -1.012$ Gs. White areas indicate there are no data points. **b** Horizontal line-cuts taken from (a). **c** $\eta$ as a function of $V_{GL}$ and $V_{GR}$ at $B = 2.755$ Gs for device B. White areas indicate there are no data points. **d** Horizontal line-cuts taken from (c). **e** Calculated $\eta$ as a function of the JJs' chemical potential $U_L$ and $U_R$ using the Furusaki-Tsukada's method with $\delta_R = 1.15\pi$. White areas indicate there are no data points. **f** Three-dimensional plot of (e).

In summary, we observed the JDE in Andreev molecules based on InAs nanowires. The
11

results reveal that the coherent coupling of two JJs enables the non-local breaking of time-reversal and spatial-inversion symmetries. By tuning both non-local phase and gate voltages, we can effectively control the JDE, with the non-local phase playing a pivotal role in reversing the sign of the diode efficiency. This behavior directly reflects the modulation of the dECT and dCAR processes, which could be essential for constructing Kitaev chains. The observed central-peak feature of the diode efficiency in the gate-voltage space further underscores the complex interplay between local and non-local effects. Our theoretical models, based on Furusaki-Tsukada's method and exact diagonalization method, provide a consistent framework that aligns with the experimental observations. These findings demonstrate the versatility of Andreev molecules as a platform for investigating non-local modulation in multi-JJ systems and offer insights into controlling superconducting phenomena in future device architectures.

## Methods

**Device fabrication.** First, Ti/Au (3/10 nm) bottom finger gates are deposited on a Si/SiO$_2$ substrate, followed by the transfer of a ~30 nm thick hBN flake onto the gates as a dielectric layer. InAs nanowires are then transferred onto the hBN. Both InAs and hBN are precisely transferred onto the bottom gates via a hot-release glue transfer platform, which is capable of picking up the nanowires or flakes at low temperature and then dropping the nanowires or flakes at high temperature. The epitaxial aluminum of InAs nanowires is etched by Transene D and the residues are removed using deionized water. Finally, Al electrodes (80 nm thick) are deposited, after Ar etching in order to remove the oxidized layer.

**Transport measurements.** The transport measurements are carried out in a dilution refrigerator at 10 mK. Keithly 2612 is used to apply DC current bias and gate voltages. Differential resistance $dV/dI$ is measured through a standard low-frequency (17.77 Hz) lock-in technique (LI5640). A 10 MΩ resistor is connected in series between the AC voltage source LI5640 and the device for current-driven measurement. The original data, i.e., the $dV/dI - I$ curves are obtained directly, and the $I - V$ curves (Fig. S4a) are obtained by the numerical integration. In order to control the small magnetic field accurately, Keithley 2400 current source is used to provide the current for the magnet.

## Acknowledgments

This work was supported by the National Key Research and Development Program of China (2022YFA1403400); by the National Natural Science Foundation of China (12074417, 92065203, 92365207, 12374459, 61974138, 92065106 and 12474049); by the Strategic Priority Research Program of Chinese Academy of Sciences (XDB33000000); by the Synergetic Extreme Condition User Facility sponsored by the National Development and Reform Commission; and by the Innovation Program for Quantum Science and Technology (2021ZD0302600 and 2021ZD0301800)). D. P. acknowledges the support from Youth Innovation Promotion Association, Chinese Academy of Sciences (2017156 and Y2021043).

## Author contributions

F.Q. conceived the project. S.Z. fabricated the devices with the help from J.H., Z.J, M.W., Y.J., J.H., and E.Z. S.Z. performed the transport measurements with the help from J.H., X.Y., and Z.L. Y.M. and B.L. carried out the theoretical calculations. D.P. and J.Z. provided InAs nanowires. X.C., B.T., Z.D., P.L., J.S., X.S., G.L., L.L. contributed to the discussion and revision of the manuscript.

## Competing interests

The authors declare no competing interests.

## Data availability

The data that support the findings of this study are available from the corresponding authors on reasonable request. Source data are provided with this paper (see Supplementary Data).

## References

1. Nadeem, M., Fuhrer, M. S. & Wang, X., The superconducting diode effect. *Nature Reviews Physics* **5**, 558-577 (2023).
2. Ando, F. *et al.*, Observation of superconducting diode effect. *Nature* **584**, 373-376 (2020).

3. Itahashi, Y. M. *et al.*, Nonreciprocal transport in gate-induced polar superconductor SrTiO3. *Science Advances* **6**, eaay 9120 (2020).

4. Narita, H. *et al.*, Field-free superconducting diode effect in noncentrosymmetric superconductor/ferromagnet multilayers. *Nature Nanotechnology* **17**, 823-828 (2022).

5. Yasuda, K. *et al.*, Nonreciprocal charge transport at topological insulator/superconductor interface. *Nature Communications* **10**, 2734 (2019).

6. Wu, Y. *et al.*, Nonreciprocal charge transport in topological kagome superconductor CsV3Sb5. *npj Quantum Materials* **7**, 105 (2022).

7. Masuko, M. *et al.*, Nonreciprocal charge transport in topological superconductor candidate Bi2Te3/PdTe2 heterostructure. *npj Quantum Materials* **7**, 104 (2022).

8. Lyu, Y.-Y. *et al.*, Superconducting diode effect via conformal-mapped nanoholes. *Nature Communications* **12**, 2703 (2021).

9. Hou, Y. *et al.*, Ubiquitous Superconducting Diode Effect in Superconductor Thin Films. *Physical Review Letters* **131**, 027001 (2023).

10. Gutfreund, A. *et al.*, Direct observation of a superconducting vortex diode. *Nature Communications* **14**, 1630 (2023).

11. Anh, L. D. *et al.*, Large superconducting diode effect in ion-beam patterned Sn-based superconductor nanowire/topological Dirac semimetal planar heterostructures. *Nature Communications* **15**, 8014 (2024).

12. Gao, A. *et al.*, An antiferromagnetic diode effect in even-layered MnBi2Te4. *Nature Electronics* **7**, 751-759 (2024).

13. Le, T. *et al.*, Superconducting diode effect and interference patterns in kagome CsV3Sb5. *Nature* **630**, 64-69 (2024).

14. Wakamura, T. *et al.*, Gate-tunable giant superconducting nonreciprocal transport in few-layer Td−MoTe2. *Physical Review Research* **6**, 013132 (2024).

15. Lin, J.-X. *et al.*, Zero-field superconducting diode effect in small-twist-angle trilayer graphene. *Nature Physics* **18**, 1221-1227 (2022).

16. de Vries, F. K. *et al.*, Gate-defined Josephson junctions in magic-angle twisted bilayer graphene. *Nature Nanotechnology* **16**, 760-763 (2021).

17. Turini, B. *et al.*, Josephson Diode Effect in High-Mobility InSb Nanoflags. *Nano Letters* **22**, 8502-8508 (2022).

18. Golod, T. & Krasnov, V. M., Demonstration of a superconducting diode-with-memory, operational at zero magnetic field with switchable nonreciprocity. *Nature Communications* **13**, 3658 (2022).

19. Bauriedl, L. *et al.*, Supercurrent diode effect and magnetochiral anisotropy in few-layer NbSe2. *Nature Communications* **13**, 4266 (2022).

20. Baumgartner, C. *et al.*, Supercurrent rectification and magnetochiral effects in symmetric Josephson junctions. *Nature Nanotechnology* **17**, 39-44 (2021).

21. Pal, B. *et al.*, Josephson diode effect from Cooper pair momentum in a topological semimetal. *Nature Physics* **18**, 1228-1233 (2022).

22. Jeon, K.-R. *et al.*, Zero-field polarity-reversible Josephson supercurrent diodes enabled by a proximity-magnetized Pt barrier. *Nature Materials* **21**, 1008-1013 (2022).

23. Chen, P. *et al.*, Edelstein Effect Induced Superconducting Diode Effect in Inversion Symmetry Breaking MoTe2 Josephson Junctions. *Advanced Functional Materials* **34**, 2311229 (2023).

24. Sundaresh, A., Väyrynen, J. I., Lyanda-Geller, Y. & Rokhinson, L. P., Diamagnetic mechanism of critical current non-reciprocity in multilayered superconductors. *Nature Communications* **14**, 1628 (2023).
25. Gupta, M. *et al.*, Gate-tunable superconducting diode effect in a three-terminal Josephson device. *Nature Communications* **14**, 3078 (2023).
26. Díez-Mérida, J. *et al.*, Symmetry-broken Josephson junctions and superconducting diodes in magic-angle twisted bilayer graphene. *Nature Communications* **14**, 2396 (2023).
27. Costa, A. *et al.*, Sign reversal of the Josephson inductance magnetochiral anisotropy and 0–π-like transitions in supercurrent diodes. *Nature Nanotechnology* **18**, 1266-1272 (2023).
28. Matsuo, S. *et al.*, Josephson diode effect derived from short-range coherent coupling. *Nature Physics* **19**, 1636-1641 (2023).
29. Trahms, M. *et al.*, Diode effect in Josephson junctions with a single magnetic atom. *Nature* **615**, 628-633 (2023).
30. Banerjee, A. *et al.*, Phase Asymmetry of Andreev Spectra from Cooper-Pair Momentum. *Physical Review Letters* **131**, 196301 (2023).
31. Wu, H. *et al.*, The field-free Josephson diode in a van der Waals heterostructure. *Nature* **604**, 653-656 (2022).
32. Yun, J. *et al.*, Magnetic proximity-induced superconducting diode effect and infinite magnetoresistance in a van der Waals heterostructure. *Physical Review Research* **5**, L022064 (2023).
33. Mazur, G. P. *et al.*, Gate-tunable Josephson diode. *Physical Review Applied* **22**, 054034 (2024).
34. Su, H. *et al.*, Microwave-Assisted Unidirectional Superconductivity in Al-InAs Nanowire-Al Junctions under Magnetic Fields. *Physical Review Letters* **133**, 087001 (2024).
35. Valentini, M. *et al.*, Parity-conserving Cooper-pair transport and ideal superconducting diode in planar germanium. *Nature Communications* **15**, 169 (2024).
36. Paolucci, F., De Simoni, G. & Giazotto, F., A gate- and flux-controlled supercurrent diode effect. *Applied Physics Letters* **122**, 042601 (2023).
37. Ciaccia, C. *et al.*, Gate-tunable Josephson diode in proximitized InAs supercurrent interferometers. *Physical Review Research* **5**, 033131 (2023).
38. Li, Y. *et al.*, Interfering Josephson diode effect in Ta2Pd3Te5 asymmetric edge interferometer. *Nature Communications* **15**, 9031 (2024).
39. Greco, A., Pichard, Q., Strambini, E. & Giazotto, F., Double loop dc-SQUID as a tunable Josephson diode. *Applied Physics Letters* **125**, 072601 (2024).
40. Matsuo, S. et al., Shapiro response of superconducting diode effect derived from Andreev molecules. Physical Review B 111, 094512 (2025).
41. Coraiola, M. et al., Flux-Tunable Josephson Diode Effect in a Hybrid Four-Terminal Josephson Junction. ACS Nano 18, 9221-9231 (2024).
42. Pillet, J. D., Benzoni, V., Griesmar, J., Smirr, J. L. & Girit, Ç. Ö., Nonlocal Josephson Effect in Andreev Molecules. *Nano Letters* **19**, 7138-7143 (2019).
43. Haxell, D. Z. *et al.*, Demonstration of the Nonlocal Josephson Effect in Andreev Molecules. *Nano Letters* **23**, 7532-7538 (2023).
44. Matsuo, S. et al., Observation of nonlocal Josephson effect on double InAs nanowires. *Communications Physics* **5**, 221 (2022).
45. Matsuo, S. *et al.*, Phase engineering of anomalous Josephson effect derived from Andreev molecules. *Science Advances* **9**, eadj3698 (2023).

46. Pillet, J. D. *et al.*, Josephson diode effect in Andreev molecules. *Physical Review Research* **5**, 033199 (2023).
47. Leijnse, M. & Flensberg, K., Parity qubits and poor man's Majorana bound states in double quantum dots. *Physical Review B* **86**, 134528 (2012).
48. Sau, J. D. & Sarma, S. D., Realizing a robust practical Majorana chain in a quantum-dot-superconductor linear array. *Nature Communications* **3**, 964 (2012).
49. Liu, C.-X., Wang, G., Dvir, T. & Wimmer, M., Tunable Superconducting Coupling of Quantum Dots via Andreev Bound States in Semiconductor-Superconductor Nanowires. *Physical Review Letters* **129**, 267701 (2022).
50. Dvir, T. *et al.*, Realization of a minimal Kitaev chain in coupled quantum dots. *Nature* **614**, 445-450 (2023).
51. Zatelli, F. *et al.*, Robust poor man's Majorana zero modes using Yu-Shiba-Rusinov states. *Nature Communications* **15**, 7933 (2024).
52. ten Haaf, S. L. D. *et al.*, A two-site Kitaev chain in a two-dimensional electron gas. *Nature* **630**, 329-334 (2024).
53. ten Haaf, S. L. D. *et al.*, Observation of edge and bulk states in a three-site Kitaev chain. *Nature* **641**, 890–895 (2025).
54. Bordin, A. *et al.*, Enhanced Majorana stability in a three-site Kitaev chain. *Nature Nanotechnology* **20**, (726-731) (2025).
55. Pan, D. *et al.*, In Situ Epitaxy of Pure Phase Ultra-Thin InAs-Al Nanowires for Quantum Devices. *Chinese Physics Letters* **39**, 058101 (2022).

# Supplementary Information for

# Josephson diode effect in nanowire-based Andreev molecules

Shang Zhu[1,2,#], Yiwen Ma[1,2,#], Jiangbo He[1,3], Xiaozhou Yang[1,2], Zhongmou Jia[1,2], Min Wei[1,2], Yiping Jiao[1,4], Jiezhong He[1,2], Enna Zhuo[1,2], Xuewei Cao[4], Bingbing Tong[1,5], Ziwei Dou[1], Peiling Li[1,5], Jie Shen[1], Xiaohui Song[1,5], Zhaozheng Lyu[1,5], Guangtong Liu[1,5], Dong Pan[6,*], Jianhua Zhao[6,7], Bo Lu[8,*], Li Lu[1,2,5,*], Fanming Qu[1,2,5,*]

[1] Beijing National Laboratory for Condensed Matter Physics, Institute of Physics, Chinese Academy of Sciences, Beijing 100190, China
[2] University of Chinese Academy of Sciences, Beijing 100049, China
[3] Key Laboratory of Low-Dimensional Quantum Structures and Quantum Control of Ministry of Education, Department of Physics and Synergetic Innovation Center of Quantum Effects and Applications, Hunan Normal University, Changsha 410081, China
[4] School of Physics, Nankai University, Tianjin 300071, China
[5] Hefei National Laboratory, Hefei 230088, China
[6] State Key Laboratory of Superlattices and Microstructures, Institute of Semiconductors, Chinese Academy of Sciences, Beijing 100083, China
[7] National Key Laboratory of Spintronics, Hangzhou International Innovation Institute, Beihang University, Hangzhou 311115, China
[8] Department of Physics, Tianjin University, Tianjin 300072, China

[#] These authors contributed equally to this work.

[*] Email: pandong@semi.ac.cn; billmarx@tju.edu.cn; lilu@iphy.ac.cn; fanmingqu@iphy.ac.cn

**Contents:**

**Supplementary Note 1. Numerical calculation using the Furusaki-Tsukada's method**

**Supplementary Note 2. Numerical calculation using the exact diagonalization method**

**Supplementary Note 3. Energy spectrum and CPR of an Andreev molecule**

**Supplementary Note 4. Junction transmission extracted by the OTBK theory**

**Supplementary Note 5. Information of device B**

**Supplementary Note 6. Non-local Josephson effect and JDE in device B**

**Supplementary Note 7. Dependence of supercurrent on local gate voltage for device B**

**Supplementary Note 1. Numerical calculation using the Furusaki-Tsukada's method**

In the following, we explain the numerical calculations of the Josephson current in the S-N-S-N-S Andreev molecule device (Fig. S1) based on the Furusaki-Tsukada's method. The Andreev molecule system can be described by the spin-degenerate B-dG Hamiltonian

$$H = \begin{bmatrix} \varepsilon - U(x) - \mu & \Delta(x) \\ \Delta^*(x) & -\varepsilon^* + U(x) + \mu \end{bmatrix}, \quad (S1)$$

in the electron-hole space with $\varepsilon$ being the kinetic energy $-\frac{\hbar^2 \partial_x^2}{2m}$ and $\mu$ being the chemical potential in the bulk state. The pair potential $\Delta(x)$ is given by

$$\Delta(x) = \begin{cases} \Delta_0 e^{i\delta_L}, & 0 < x < L_1 \\ 0, & L_1 < x < L_2 \\ \Delta_0, & L_2 < x < L_3 \\ 0, & L_3 < x < L_4 \\ \Delta_0 e^{i\delta_R}, & L_4 < x < L_5 \end{cases}, \quad (S2)$$

where $\delta_L(\delta_R)$ is the macroscopic superconducting phase on the left (right) superconductor. The variation of the chemical potential of our setup is modeled by $U(x) = U_S \Theta(L_1 - x) + U_L(x)\Theta(x - L_1)\Theta(L_2 - x) + U_S\Theta(x - L_2)\Theta(L_3 - x) + U_R(x)\Theta(x - L_3)\Theta(L_4 - x) + U_S\Theta(x - L_4)$ where $U_L$ and $U_R$ can be tuned by the left and right gate voltage in the normal region, respectively. We set $U_S = 0$ and solve the B-dG wave function of the above Hamiltonian. For example, we consider an incident electron-like quasiparticle with energy $E$ from the left superconductor and the corresponding wave function for the left superconductor is written as

$$\Psi^e(x < 0) = \begin{bmatrix} ue^{i\varphi/2} \\ ve^{-i\varphi/2} \end{bmatrix} e^{ip^+ x} + a_e \begin{bmatrix} ve^{i\varphi/2} \\ ue^{-i\varphi/2} \end{bmatrix} e^{ip^- x} + b_e \begin{bmatrix} ue^{i\varphi/2} \\ ve^{-i\varphi/2} \end{bmatrix} e^{-ip^+ x}. \quad (S3)$$

$p^+$ and $p^-$ are the wave vectors and they are $p^\pm = \sqrt{2m(\mu \pm \sqrt{E^2 - \Delta^2})}/\hbar$. The factors $u$ and $v$ are given by $u(v) = \sqrt{(1 \pm \sqrt{E^2 - \Delta^2/E})/2}$. As a hole-like quasiparticle injects from the left side, we use the wave function $\Psi^h(x)$ for the left superconductor

$$\Psi^h(x < 0) = \begin{bmatrix} ue^{i\varphi/2} \\ ve^{-i\varphi/2} \end{bmatrix} e^{-ip^- x} + a_e \begin{bmatrix} ve^{i\varphi/2} \\ ue^{-i\varphi/2} \end{bmatrix} e^{-ip^+ x} + b_e \begin{bmatrix} ue^{i\varphi/2} \\ ve^{-i\varphi/2} \end{bmatrix} e^{ip^- x}. \quad (S4)$$

The wave functions for other regions are obtained in the same way. Then the coefficients $a_{e(h)}$ and $b_{e(h)}$ can be determined by matching wave functions at each boundary,

$$\Psi(x^+) = \Psi(x^-), \quad (S5)$$
$$\partial_x \Psi(x^+) = \partial_x \Psi(x^-), \quad (S6)$$

We need $a_e$ and $a_h$ to calculate the Josephson current via Furusaki-Tsukada's formula[1]

$$I_L = \frac{ekT\Delta}{\hbar} \sum_{\omega_n} \frac{p^+ + p^-}{\sqrt{\omega_n^2 + \Delta^2}} \left[\frac{a_e}{p^+} - \frac{a_h}{p^-}\right]. \quad (S7)$$

Here, we have made analytical continuation of incident quasiparticle energy $E \to i\omega_n$ into Matsubara frequencies $\omega_n = \pi K_B T(2n + 1), (n = 0, \pm 1, \pm 2 \cdots)$ so that $a_e$ and $p^\pm$ have dependency on $\omega_n$. We further adopt the BCS relation for its temperature dependence:

$\Delta(T) = \Delta_0 \tanh(1.74\sqrt{T_c/T - 1})$ with $\Delta_0 = 1.76 K_B T_c$, $T_c$ is the critical temperature. Thus, the Josephson current and the diode efficiency can be obtained. Figs. 4c and 4d are calculated with parameters: $N_2 = N_4 = 1600$, $N_3 = 3000$, $\delta_R = 1.15\pi$.

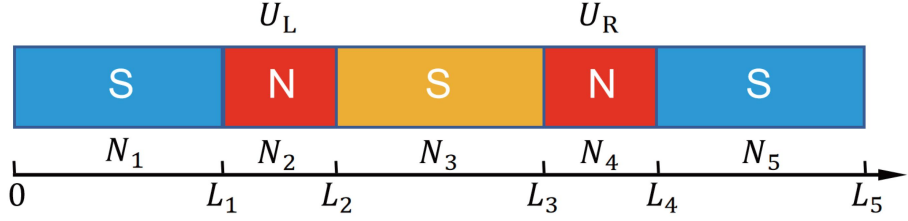

**Fig. S1. S-N-S-N-S model of an Andreev molecule.** The lengths of the S-N-S-N-S sections are defined as $N_1$ - $N_5$, and the corresponding boundaries are defined as $L_1$ - $L_5$.

**Supplementary Note 2. Numerical calculation using the exact diagonalization method**

To investigate the energy spectrum of an Andreev molecule, it is convenient to adopt an exact diagonalization method using a tight-binding model. In a S-N-S-N-S device, the length of each section is shown in Fig. S1. The Hamiltonian is

$$H = \sum_{i,s} U_i c_{i,s}^\dagger c_{i,s} - \sum_{<i,i+1>,s} t\left(c_{i,s}^\dagger c_{i+1,s} + h.c.\right) + \sum_i \left(\Delta_i c_{i,\uparrow}^\dagger c_{i,\downarrow}^\dagger + \Delta_i^* c_{i,\downarrow} c_{i,\uparrow}\right),\qquad(S8)$$

where

$$U_i = \begin{cases} U_L, & L_1 < i < L_2 \\ U_R, & L_3 < i < L_4, \\ 0, & \text{else} \end{cases} \qquad(S9)$$

$$\Delta_i = \begin{cases} \Delta_0 e^{i\delta_L}, & 0 < i < L_1 \\ \Delta_0, & L_2 < i < L_3 \\ \Delta_0 e^{i\delta_R}, & L_4 < i < L_5 \\ 0, & \text{else} \end{cases} \qquad(S10)$$

$c_{is}$ and $c_{is}^\dagger$ are the annihilation and creation operators of electrons, $U_i$ is the on-site potential, and $\Delta_i$ is the superconducting order parameters on each site. We assume that the on-site potential only locates at the left and right junction region with height of $U_L$ and $U_R$, respectively, which determine the transmission of the JJs. The phases of the left and right JJs are introduced by setting the superconducting order parameters of each lead.

For a given $\delta_R$ determined by the magnetic flux through the superconducting loop, the energy spectrum includes the continuum states and the discrete ABSs inside the gap, obtained by the exact diagonalization method. The Josephson current through the left junction is obtained by calculating the derivative of the energy $E_n$ with respect to the phase $\delta_L$,

$$I_L(\delta_L; \delta_R) = \sum_{E_n < 0} \frac{\partial E_n(\delta_L; \delta_R)}{\partial \delta_L}. \qquad(S11)$$

Thus, the energy spectrum (Figs. 1d and S2a) and the diode efficiency (Figs. S3a and S3b) can be calculated using this exact diagonalization method.

## Supplementary Note 3. Energy spectrum and CPR of an Andreev molecule

Figures S2a and b show the energy spectrum and asymmetric CPR calculated by exact diagonalization method at $\delta_R = 1.15\pi$, respectively. At $\delta_L = \delta_L^+$, both the ABSs and the continuum states (at $E < 0$) contribute the same-direction supercurrent to $I_{Lc}^+$, where dECT is the dominant mechanism. In contrast, at $\delta_L = \delta_L^-$, the supercurrent direction carried by the two ABSs is opposite, i.e., dCAR is dominant, resulting in a largely reduced critical supercurrent $|I_{Lc}^-|$. In addition, the contribution of the continuum states is also smaller at $\delta_L = \delta_L^-$ than at $\delta_L = \delta_L^+$, further enhancing the JDE. Therefore $I_{Lc}^+ > |I_{Lc}^-| \rightarrow \eta > 0$, as shown in Fig. S2b. Note that the scenario is reversed compared to that at $\delta_R = 0.85\pi$, as explained in the main text.

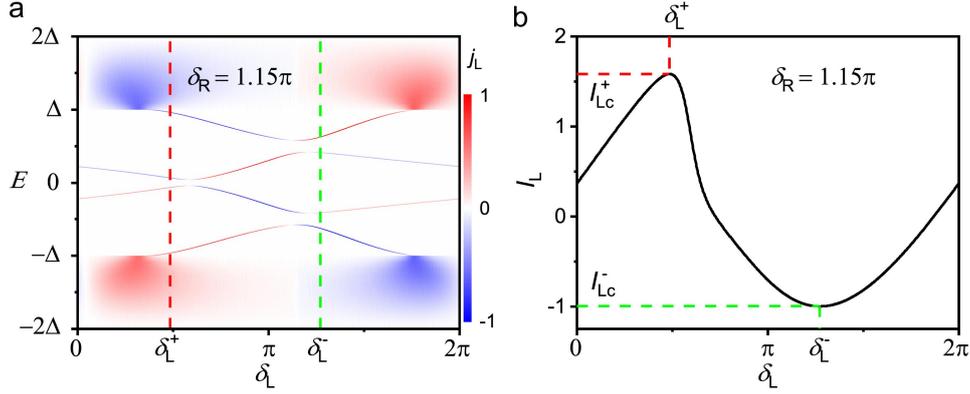

**Fig. S2. Energy spectrum and CPR of an Andreev molecule. a** Energy spectrum of the Andreev molecule at $\delta_R = 1.15\pi$. For $|E| < \Delta$, the red and blue color of the lines indicate the positive and negative supercurrent of the left JJ carried by the ABSs, respectively. For $|E| \geq \Delta$, the color corresponds to the supercurrent density $j_L$ of the continuum states, as indicated by the color bar. **b** CPR of the left JJ at $\delta_R = 1.15\pi$. The red and green dashed lines in (a), (b) mark the positions of $I_{Lc}^+$ at $\delta_L = \delta_L^+$ and $I_{Lc}^-$ at $\delta_L = \delta_L^-$, respectively. Parameters used for calculations: $t = 1$, $\Delta_0 = 0.2$, $N_1 = N_5 = 200$, $N_2 = N_4 = 1$, $N_3 = 10$, the same as Figs. 1d and 1e.

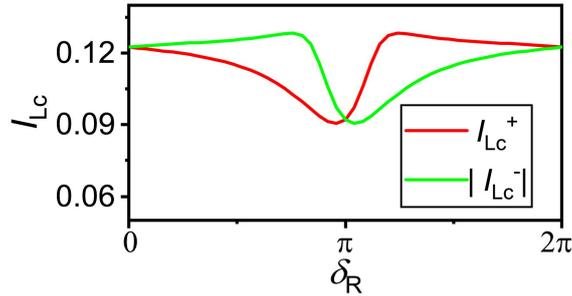

**Fig. S3. Critical supercurrent $I_{Lc}^+$ and $|I_{Lc}^-|$ as a function of $\delta_R$.** When $0 < \delta_R < \pi$, $I_{Lc}^+ < |I_{Lc}^-|$; while $\pi < \delta_R < 2\pi$, $I_{Lc}^+ > |I_{Lc}^-|$.

Figure S4 shows Andreev molecule energy spectrum with different $\delta_R$. When $\delta_R = 0.3\pi$ or $0.6\pi$, $I_{Lc}^+ < |I_{Lc}^-|$ (Fig. S3), dECT dominates at $\delta_L = \delta_L^-$ while dCAR dominates at $\delta_L = \delta_L^+$. As for $\delta_R = 1.4\pi$ or $1.7\pi$, $I_{Lc}^+ > |I_{Lc}^-|$ (Fig. S3), dECT dominates at $\delta_L = \delta_L^+$ while dCAR dominates at $\delta_L = \delta_L^-$.

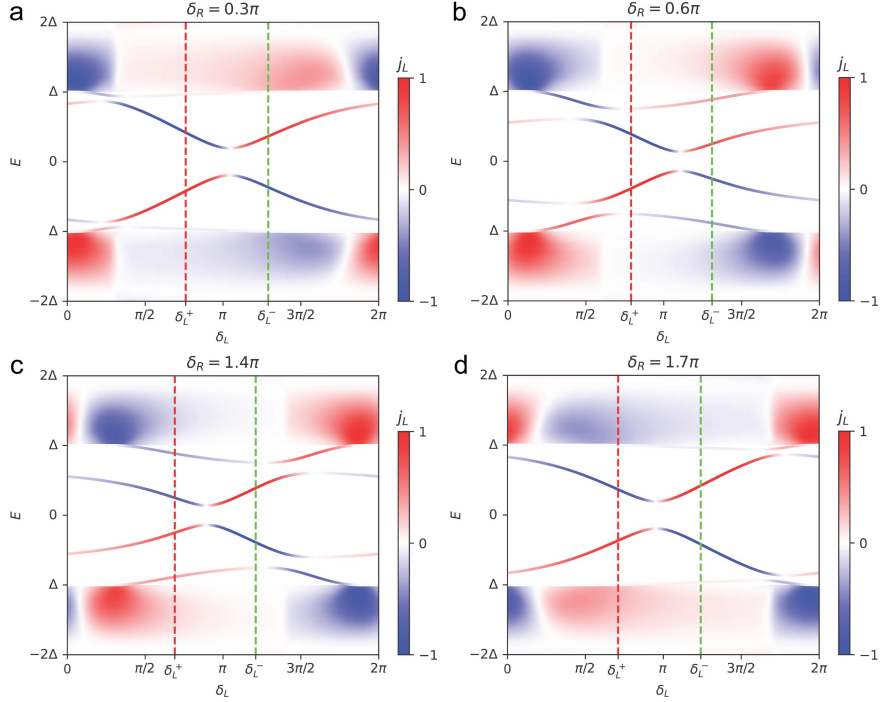

**Fig. S4. Andreev molecule energy spectrum with different $\delta_R$. a, b, c, d** Energy spectrum of the Andreev molecule at $\delta_R = 0.3\pi, 0.6\pi, 1.4\pi$ and $1.7\pi$. The red and green dashed lines mark the positions of $\delta_L^+$ and $\delta_L^-$, respectively.

Next, we can get $\eta$ as a function of $U_L$ and $U_R$, as shown in Figs. S5a and b, in the form of two-dimensional and three-dimensional graphs, respectively. It can be seen that $\eta$ presents a central-peak feature near $U_L = U_R = 0$. These results are qualitatively consistent with our experimental observations and the results calculated by the Furusaki-Tsukada's method.

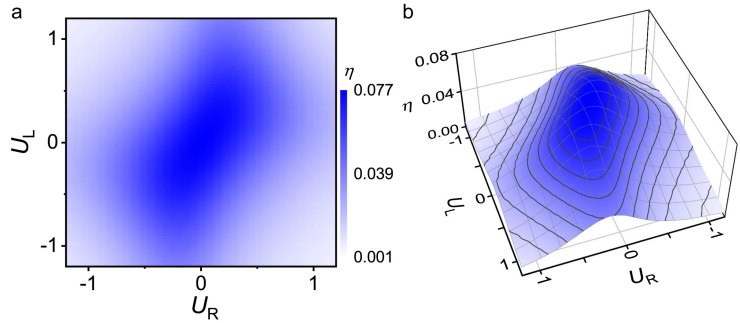

**Fig. S5. Numerical calculation results using the exact diagonalization method. a** $\eta$ as a function of the JJs' chemical potential $U_L$ and $U_R$. **b** Three-dimensional plot of (a). Parameters used for calculations: $t = 1$, $\Delta_0 = 0.2$, $K_B T = 0.015$, $N_1 = N_5 = 50$, $N_2 = N_4 = 1$, $N_3 = 10$, $\delta_R = 1.15\pi$.

**Supplementary Note 4. Junction transmission extracted by the OTBK theory**

The *I-V* curve of the left JJ of device A can be obtained by integrating the measured $dV/dI$ vs. $I$ curve, as shown in Fig. S6a. When the JJ enters the normal state (far above the energy gap), the *I-V* curve is approximately linear. The excess current $I_{exc} = 17.6$ nA can be extracted by extending the linear part, as indicated by the red dashed line whose slope corresponds to the normal state resistance $R_N = 3950\ \Omega$. According to OTBK theory, the barrier strength $Z$ of the interface is incorporated by the following equation[2]:

$$\frac{eI_{exc}R_N}{\Delta} = 2(1+2Z^2)\times\tanh^{-1}(2Z\sqrt{\frac{(1+Z^2)}{(1+6Z^2+4Z^4)}})\times$$
$$(Z\sqrt{(1+Z^2)(1+6Z^2+4Z^4)})^{-1}-\frac{4}{3}, \quad (S12)$$

where the superconducting gap $\Delta = 190\ \mu eV$. Figure S4b shows the curve corresponding to Eq. (S12), and for the left JJ $Z = 0.947$. Therefore, the transmission of the left JJ $\tau = 1/(1+Z^2) \approx 0.53$. Note that this is a rather rough estimate of the transmission.

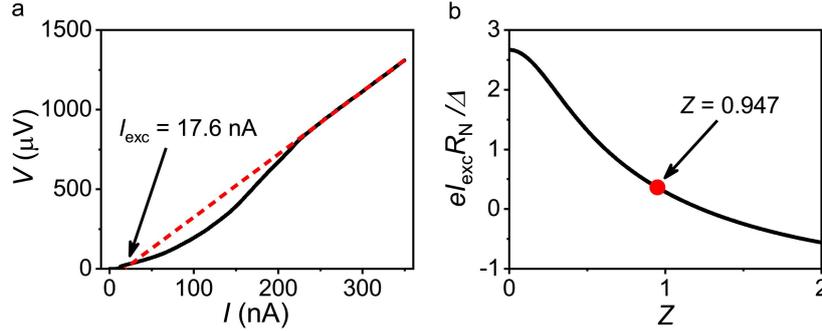

**Fig. S6. Junction transmission extracted by the OTBK theory. a** *I-V* curve of the left JJ of device A, and the excess current $I_{exc} = 17.6$ nA. **b** $eI_{exc}R_N/\Delta$ as a function of the barrier strength $Z$ according to the OTBK theory, and for the left JJ $Z = 0.947$.

**Supplementary Note 5. Information of device B**

Figure S7 shows the false-color SEM image of device B. The fabrication process of device B is essentially the same as device A. The length of the two JJs is about 70 nm each. Thin Al with a length about $l = 300$ nm is retained between the two JJs.

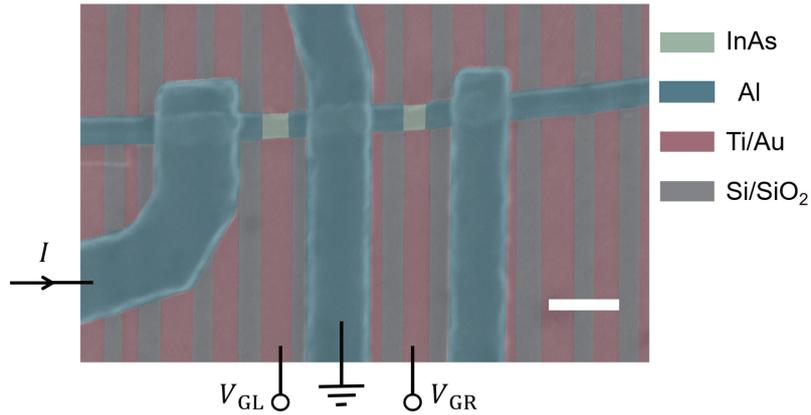

**Fig. S7. False-color SEM image of device B.** The scale bar is 200 nm. An Al loop is fabricated for the right JJ (not shown). The hBN dielectric layer sits between the Ti/Au finger gates and the nanowire, but is transparent in this image.

**Supplementary Note 6. Non-local Josephson effect and JDE in device B**

Figures S8a and S8b show the differential resistance $dV/dI$ of the left JJ of device B as a function of $I$ and $B$, at fixed $V_{GL} = V_{GR} = 0$ V, scanned towards the positive and negative current direction, respectively. Both the positive and negative critical supercurrent show periodic oscillations, demonstrating the non-local Josephson effect. The critical supercurrent $I_{Lc}^+$ and $|I_{Lc}^-|$ can be extracted from Figs. S8a and S8b, as plotted in Fig. S8c. Similar to Fig. 2b for device A, within a certain range of $B$, $I_{Lc}^+ \neq |I_{Lc}^-|$, which is the JDE. The $\eta$ vs. $B$ curve can then be obtained, as shown in Fig. S8d. Compared to device A, the critical supercurrent of device B has a smaller oscillation amplitude and a lower diode efficiency, which may be related to a weaker coupling between the two JJs due to a larger separation ($l = $ 300 nm for device B, and 200 nm for device A).

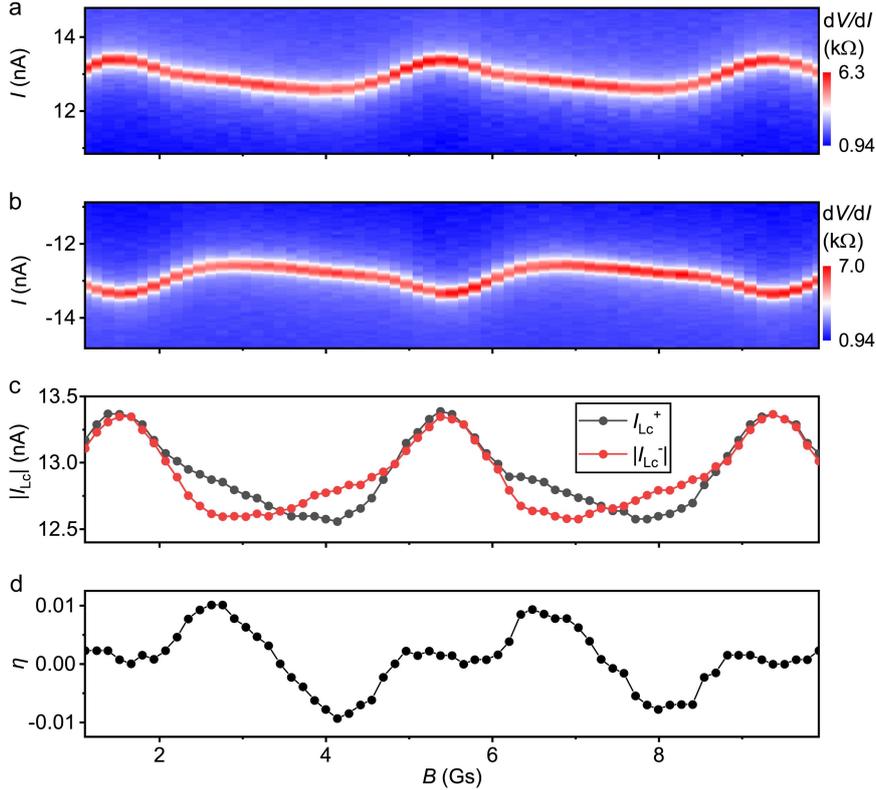

**Fig. S8. The JDE observed in device B. a, b** The differential resistance $dV/dI$ of the left JJ as a function of $I$ and $B$ with $V_{GL} = V_{GR} = 0$ V, scanned towards the positive and negative current direction, respectively. **c** $|I_{Lc}|$ as a function of $B$. **d** Diode efficiency $\eta$ as a function of $B$.

**Supplementary Note 7. Dependence of supercurrent on local gate voltage for device B**

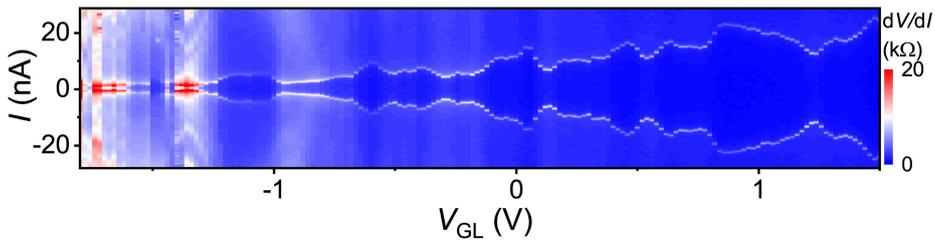

**Fig. S9. Dependence of supercurrent on local gate voltage for device B.** The differential resistance $dV/dI$ of left JJ as a function of $I$ and $V_{GL}$. The critical supercurrent decreases as $V_{GL}$ decreases, and completely disappears when $V_{GL} = -1.8$ V.

## Supplementary References

1. Furusaki, A. & Tsukada, M., Dc josephson effect and andreev reflection. *Solid State Communications* 78, 4 (1991).

2. Niebler, G., Cuniberti, G. & Novotný, T., Analytical calculation of the excess current in the Octavio‑Tinkham‑Blonder‑Klapwijk theory. *Superconductor Science and Technology* 22, 085016 (2009).